**Yekimov Sergiy**
Department of Trade and Finance
Faculty of Economics and Management
Czech University of Life Sciences Prague
Kamycka 129
165 00 Praha – Suchdol, Czech Republic
https://orcid.org/0000-0001-6575-2623

# Dynamic gravity model of urban passenger transportation

## Abstract

Modeling of urban traffic flows is required due to the complexity of their successful forecasting, as well as due to the impact of various random factors on them, and the complexity of transport systems in modern cities.

Forecasting of traffic flows can be used as an effective tool for making scientifically sound decisions when creating a long-term plan for the integrated development of the city.

The complex nature of the behavior of people and transport systems requires taking into account many different external and internal factors in the process of building a realistic model, as well as using large computing resources and large amounts of source data.

In the process of planning the development of transport systems, the correct compilation of correspondence matrices, in which various modeling zones act as columns and rows, is of paramount importance.

At the same time , the elements of the correspondence matrices are not static in general , but are subject to temporary changes , primarily diurnal and seasonal . The dynamic gravitational model proposed by us in this paper is based on a system of first - order differential equations . The solution of which will allow, according to the authors, to more accurately take into account the characteristics of urban traffic flows.

If there is a set of initial data and the corresponding software, calculations of the model we offer can be made within a few hours.

## Introduction

A modern city requires a constant increase in traffic volumes, improvement of their quality, level of reliability and safety. This , in turn , requires the involvement of public and private investments in the transport industry . To reduce the risk of investing in transport infrastructure, it is necessary to study the dynamics of the development of transport infrastructure, to study the level of its congestion of all its sections.

According to El Essaili, Ali & PALAIOS, ALEXANDROS. (2020). the lack of effective approaches to traffic flow management causes an increase in accidents on the roads, the formation of traffic jams, underloading or overloading of individual sections of the transport network..

According to van Putten, Bart-Jan & Wolfe, Shawn & Dignum, Virginia. (2008) for effective traffic flow management of a modern city, a large number of different parameters characterizing traffic flows should be taken into account, taking into account the impact of various internal and external factors affecting the intensity of traffic flows.

According to White, M.J.. (1992), the use of insufficiently reliable data obtained on the basis of one-time surveys and local transport mobility data becomes the reason for creating transport models that cause inefficient management decisions.

According to Hamdan, Sadeque & Jouini, Oualid & Cheaitou, Ali & Jemai, Zied & Granberg, Tobias. (2022), in order to create an optimally balanced Master Plan for the development of a modern city, it is necessary to have an instrument for evaluating urban planning measures related to the development of transport infrastructure and large residential areas.

According to Beygi, Shervin & Bromberg, Emily & Elliott, Matthew & Krishna, Shubh & Lewis, Taryn & Schultz, Laura & Aviation, Metron & Wetherly, James & Sud, Ved. (2012), in order to study the demand for transport services, it is necessary to find out the patterns in the population's expectations based on sociological research and official statistics.

According to Sridhar, Banavar. (2009), the user of the transport network chooses the routes of their movement in the most optimal way for them corresponding to the minimum amount of costs.

According to Rios, Joseph & Ross, Kevin. (2008) the increase in the standard of living in modern megacities, the increase in the number of motor transport complicates the assessment of the need for transport services by the population.

The authors Dearmon, James. (1993) identify the following factors of the demand of city residents for transport services (Figure 1):

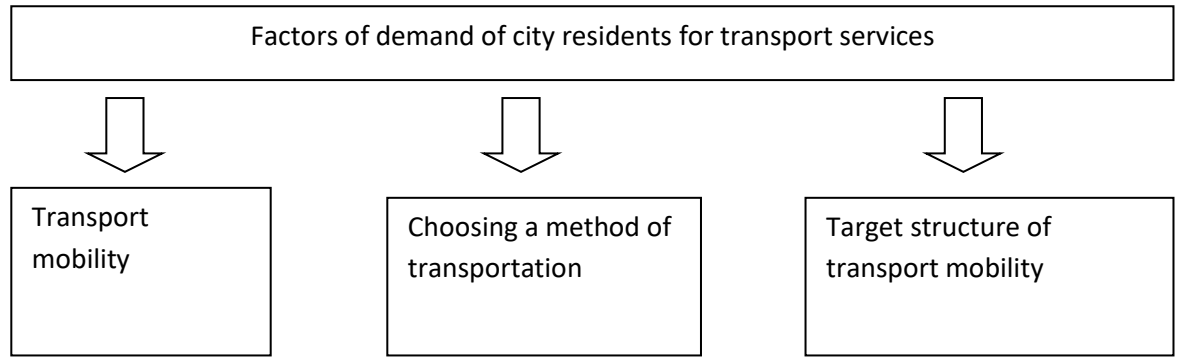

Figure 1. Factors of demand of city residents for transport services

According to Saraf, Aditya & Hunter, George & Ramamoorthy, Krishnakumar & Nagle, Gaurav & Cheng, Kevin. (2013) ; Pipa, Martin & Scerba, Marek & Mesina, J. & Wang, J.H.. (2012) ; Grabbe, Shon & Sridhar, Banavar & Mukherjee, Avijit. (2009); Abeysooriya, Milanka & Kulasekere, Chulantha. (2006). , the characteristics of the transport behavior of the city's population are necessary to create the necessary long- and medium-term forecasts for the effective development of transport infrastructure.

According to Abeysooriya, Milanka & Kulasekere, Chulantha. (2006) ; S.Yekimov & V.A.Tkachenko & K.V.Zavgorodniy & I.P.Oleshkevich & N.Sirenko (2021) the study of the population's demand for transport services makes it possible to use vehicles more optimally, increase the level of efficiency of passenger transportation management in the city and more fully satisfy the need of the population to travel around the city.

According to Rumánková, Lenka & Sanova, Petra & Kolář, P.. (2019) ; Dvořák, Marek & Smutka, Lubos & Krajčírová, Renáta & Kadekova, Zdenka & Pulkrábek, Josef. (2021) urban passenger transportation is of great social importance, therefore it is important to optimize the schedules of urban public transport and be able to correctly predict traffic flows.

According to Toušek, Zdenek & Hinke, Jana & Malinská, Barbora & Prokop, Martin. (2022) ; Krivko, Mikhail & Moravec, Lukas & Šálková, Daniela & Smutka, Lubos & Kukalová, Gabriela. (2021), the analysis of the urban transport network requires solutions to the following problems (Figure 2):

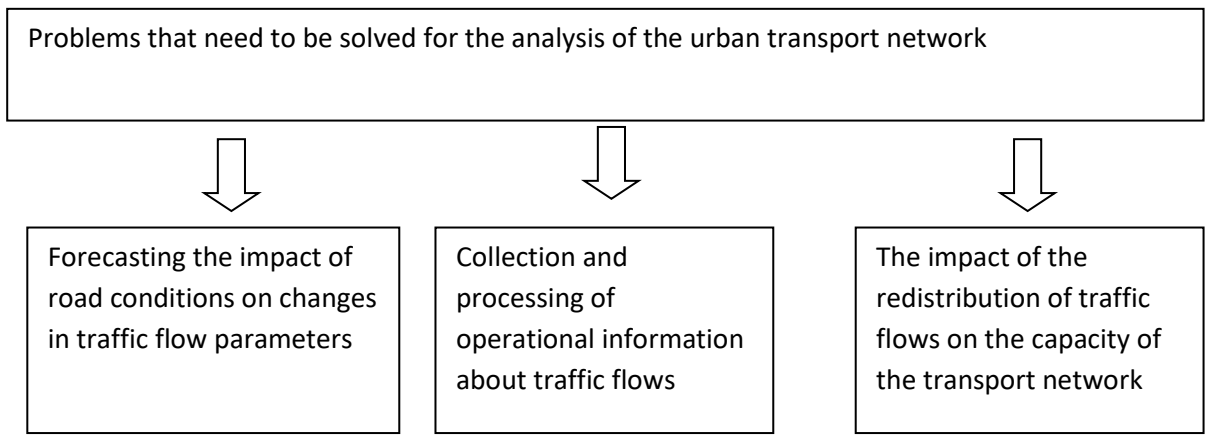

Figure 2 Problems that need to be solved for the analysis of the urban transport network

According to Krivko, Mikhail & Moravec, Lukas & Kukalová, Gabriela & Smutka, Lubos & Šálková, Daniela. (2021) ; Šálková, Daniela & Kučera, Petr & Moravec, Lukas. (2017), there are many approaches to predicting the level of intensity of traffic flows. Which can be divided into two groups :

1) Methods based on the application of growth coefficients. It is based on growth coefficients determined based on data on the current state of traffic flows and extrapolation of data for past periods.

2) A method based on the construction of models of traffic flows. It is based on the study of the goals and reasons for the movement of people using road transport.

Linear, entropic, and gravitational models are used for the mathematical description of urban traffic flows Caves, D.W.& Christensen, L.R.& Diewert, W.E. (1982)

The linear model describes outgoing and incoming traffic flows as changing linearly over time. The intensity of the traffic flow is characterized by the number of correspondences that relate to the *i*-th fragment of the transport network

$$P_{i,j} = \mu_i \rho_{i,j} \qquad (1)$$

where $P_{i,j}$ - future transport correspondence , $-\rho_{i,j}$ transport orrespondence at the present time , $\mu_i$ - growth coefficient , depending on the number of vehicles , the number of population , income per population , business activity .

The gravitational model describes incoming and outgoing traffic flows by analogy with the formula of the law of universal gravitation.

$$\rho_{i,j} = \frac{s_i d_j}{c_{i,j}} \tag{2}$$

Where $s_i$ - number of vehicles leaving the source $i, i \in S$, $d_j$ - number of vehicles entering from the source $j, j \in D$, $c_{i,j}$ – utility function corresponding to moving to the drain $j$ from the source $i$.

Expression (2) must meet the following additional conditions:

$$\sum_{i \in S} s_i = \sum_{j \in D} d_j = R, \sum_{j=1}^{n} \rho_{i,j} = s_i, \sum_{i=1}^{m} \rho_{ij} = d_j, \rho_{i,j} \geq 0, i \in S, j \in D \tag{3}$$

Within the entropy model, the number $\rho_{i,j}$ of correspondences between sources $i$ and sinks $j$ is determined as the maximum of the probability $P(\rho)$ function of the system state based on the correspondence matrix $\rho_{i,j}; i \in S, D \in j$.

Model (3) is subject to restrictions on the costs associated with movement within the transport system.

$$\sum_{i=1}^{n} \sum_{j=1}^{m} c_{i,j} \rho_{i,j} = C_{max}$$

In [14], a gravitational model was proposed

$$\alpha_0^s + \sum_i^I \alpha_i^s \ln Y_i^s + \sum_n^N \beta_n^s \ln X_n^s + \frac{1}{2} \sum_i^I \sum_j^J \alpha_{ij} \ln Y_i^s \ln Y_j^s +$$

$$+ \frac{1}{2} \sum_n^N \sum_m^M \beta_{nm} \ln X_n^s \ln X_m^s + \sum_i^I \sum_n^N \pi_{in} \ln Y_i^s \ln X_n^s = 1 \tag{4}$$

where $X_n^s$ - vector-input, $Y_i^s$ - vector-output, $\alpha_0^s, \alpha_i^s, \beta_n^s, \alpha_{ij}, \beta_{nm}, \pi_{in}$ - parameters that are constants

$$\alpha_{ij} = \alpha_{ji}, \quad \beta_{ij} = \beta_{ji}$$

$\sum_n^N \beta_{nm} = 0$ for $m = 1, ..., M$   $\sum_n^I \pi_{in} = 0$ for $n = 1, ..., N$

$\sum_n^N \pi_{in} = 0$ for $i = 1, ..., I$

It should be noted that the models (1),(2), (4) they are static, this does not allow us to properly take into account the dynamics of the development of the transport network.

The authors Čechura L.& Žáková Kroupová Z.& Lekešová M. (2022);Čechura, Lukáš & Zdeňka Žáková Kroupová & Antonella Samoggia (2021);Cechura, Lukas & Hockmann, Heinrich & Malý, Michal & Zakova Kroupova, Zdenka. (2015);Cechura L.& Kroupova Z.& Rudinskaya T. (2015);Čechura, L.& Žáková Kroupová, Z.& Kostlivý, V. & Lekešová, M. (2021); Cábelková, I.& Smutka, L.&

Rotterova, S.& Zhytna, O.& Kluger, V.& Mareš, D. (2022) use a gravitational TREE model M-output (y) , J-input (x)

$$\ln D_{it}^I = \alpha_0 + \sum_{m=1}^{M} \alpha_m \ln y_{m,it} + \frac{1}{2}\sum_{m=1}^{M}\sum_{n=1}^{N} \alpha_{mn} \ln y_{m,it} \ln y_{n,it} + \sum_{m=1}^{M}\sum_{j=1}^{J} \gamma_{mj} \ln y_{m,it} \ln x_{j,it} +$$

$$\sum_{j=1}^{J} \beta_j \ln x_{j,it} + \frac{1}{2}\sum_{j=1}^{J}\sum_{k=1}^{K} \beta_{jk} \ln x_{j,it} \ln x_{k,it} + \delta_t t + \frac{1}{2}\delta_{tt} t^2 + \sum_{m=1}^{M} \delta_{mt} \ln y_{m,it} t + \sum_{j=1}^{J} \delta_{jt} \ln x_{j,it} t$$

(5)

where $\alpha_{mn}, \beta_{jk}, \gamma_{mj}, \delta_{mt}, \delta_{jt}, \alpha_0, \beta_j, \delta_t, \alpha_m$ – parameters that are constants $\alpha_{ij} = \alpha_{ji}$ , $\beta_{ij} = \beta_{ji}$ $\sum_{j=1}^{J}\beta_j = 1;\ \sum_{j=1}^{J}\beta_{jk} = 0;\ \sum_{j=1}^{J}\gamma_{mj} = 0;\ \sum_{j=1}^{J}\delta_{jt} = 0$

Model (5) can be used to analyze traffic flows, but the dynamics of changes in traffic flows is approximately described by a second-order polynomial in time

$$\delta_t t + \frac{1}{2}\delta_{tt} t^2 + \sum_{m=1}^{M}\delta_{mt}\ln y_{m,it} t + \sum_{j=1}^{J}\delta_{jt}\ln x_{j,it} t \quad (6)$$

At the same time, this dependence may be more complex, which in turn may negatively affect the quality of prognosis based on such a model.

## Methods

During the implementation of this scientific work, we used an analytical method, which allowed us to consider the problems studied in the article in their unity and development.

Taking into account the objectives and goals of the research, we used a functional-structural method of scientific cognition.

As a result, the authors were able to consider a number of problems with the dynamic gravity model of urban passenger transportation

## Results

We propose a more generalized gravitational model

$$\ln D_{it}^I = \alpha_0 + \sum_{m=1}^{M} \alpha_m(t)\ln y_{m,it} + \frac{1}{2}\sum_{m=1}^{M}\sum_{n=1}^{N} \alpha_{mn}(t)\ln y_{m,it}\ln y_{n,it} +$$

$$+ \sum_{m=1}^{M}\sum_{j=1}^{J} \gamma_{mj}(t)\ln y_{m,it}\ln x_{j,it} + \sum_{j=1}^{J}\beta_j(t)\ln x_{j,it} + \frac{1}{2}\sum_{j=1}^{J}\sum_{k=1}^{K}\beta_{jk}(t)\ln x_{j,it}\ln x_{k,it}$$

<div style="text-align: right">(7)</div>

где, $\alpha_m = \alpha_m(t)$ $\alpha_{mn} = \alpha_{mn}(t)$ $\gamma_{mj} = \gamma_{mj}(t)$ $\beta_{jk} = \beta_{jk}(t)$ $\beta_j = \beta_j(t)$

Differentiate equation (7)

$$\frac{\partial D_{it}^I}{\partial \alpha_m(t)} = D_{it}^I * \ln y_{m,it} \qquad \frac{\partial D_{it}^I}{\partial \alpha_{mn}(t)} = D_{it}^I * \frac{1}{2} * \ln y_{m,it} * \ln y_{n,it}$$

$$\frac{\partial D_{it}^I}{\partial \beta_j(t)} = D_{it}^I * \ln x_{j,it} \qquad \frac{\partial D_{it}^I}{\partial \beta_{jk}(t)} = D_{it}^I * \frac{1}{2} * \ln x_{j,it} * \ln x_{k,it}$$

$$\frac{\partial D_{it}^I}{\partial \gamma_{mj}(t)} = D_{it}^I * \ln y_{m,it} * \ln x_{j,it}$$

$$\frac{dD_{it}^I}{dt} = D_{it}^I \left( \ln y_{m,it} * \frac{d\alpha_m(t)}{dt} + \frac{1}{2} * \ln y_{m,it} * \ln y_{n,it} * \frac{d\alpha_{mn}(t)}{dt} + \frac{1}{2} * \ln x_{j,it} * \ln x_{k,it} * \frac{d\beta_{jk}(t)}{dt} \right) +$$

$$+ D_{it}^I \left( \ln y_{m,it} * \ln x_{j,it} * \frac{d\gamma_{mj}(t)}{dt} + \ln x_{j,it} * \frac{d\beta_j(t)}{dt} \right)$$

After a simple transformation, we finally get

$$\frac{d(\ln(D_{it}^I))}{dt} = \ln y_{m,it} * \frac{d\alpha_m(t)}{dt} + \frac{1}{2} * \ln y_{m,it} * \ln y_{n,it} * \frac{d\alpha_{mn}(t)}{dt} + \frac{1}{2} * \ln x_{j,it} * \ln x_{k,it} * \frac{d\beta_{jk}(t)}{dt} +$$

$$+ \ln y_{m,it} * \ln x_{j,it} * \frac{d\gamma_{mj}(t)}{dt} + \ln x_{j,it} * \frac{d\beta_j(t)}{dt} \qquad (8)$$

Unlike models (1), (2), (5) model (8) contains the rate of change $\ln(D_{it}^I)$, as well as parameters $\alpha_m, \alpha_{mn}, \beta_j, \beta_{jk}, \gamma_{mj}$. This in turn allows you to take into account changes in parameters $\alpha_m, \alpha_{mn}, \beta_j, \beta_{jk}, \gamma_{mj}$ in time, what is in the models (1), (2), (5) not provided.

## Discussion

In the process of planning the development of transport systems, the correct compilation of correspondence matrices is important, where various modeling zones act as columns and rows. At the same time, the elements of the correspondence matrices are not static in general, they are subject to temporary changes and, first of all, seasonal. The dynamic gravitational model proposed by us in this paper is based on a system of first-order differential equations. The solution of which will allow, according to the authors, to more accurately take into account the characteristics of urban traffic flows.

If there is a set of source data and the appropriate software, calculations of the model we offer can be made within a few hours.

## Conclusions

Modeling of urban traffic flows is necessary due to the complexity of their forecasting, the impact of various random factors on them, as well as the complexity of transport systems in modern cities.

Forecasting of traffic flows can be used as an effective tool for making informed decisions when planning the integrated development of the city.

The complex nature of the behavior of people and transport systems requires taking into account many different factors to build a realistic model, using large computing resources and large amounts of source data.